# Ultrasensitive infrared-to-visible artificial vision via self-evolving projection guided by single-pixel detection

Yao Wang[1], Baolei Liu[2,3*], Muchen Zhu[1], Linjun Zhai[1], Dajing Wang[1], Zhaohua Yang[2,3], and Fan Wang[1*]

[1]School of Physics, Beihang University, Beijing, 100191, China
[2]School of Instrumentation and Optoelectronics Engineering, Beihang University, Beijing, 100191, China
[3] Hangzhou International Innovation Institute, Beihang University, Hangzhou, 311115, China
These authors contributed equally to this work: Yao Wang and Baolei Liu
*Correspondence: liubaolei@buaa.edu.cn; fanwang@buaa.edu.cn

**Keywords:** infrared-to-visible upconversion, ultra-low photon sensing, single-pixel detection, adaptive illumination

**Abstract:** Infrared detection and visualization are essential for augmenting human perception across diverse fields, ranging from night vision to industrial inspection and bio-imaging. Conventional infrared cameras are often hindered by high cost, bulky architecture, and complex fabrication requirements. Upconversion sensing systems offer a pixel-free and cost-effective alternative solution by upconverting low-energy infrared photons into visible-light signals. However, existing upconversion systems suffer from limitations such as high operating voltages, low quantum efficiency, and the need for intense excitation, which prevent their applications in photon-starved environments, particularly for imaging at the single- or sub-photon conditions. Here, we report self-evolving infrared-to-visible upconversion with single-pixel detection (SIVIS) that enables real-time upconverted visualization under photon-starved conditions by integrating self-evolving projection with single-pixel sensing. SIVIS iteratively optimizes illumination patterns with a digital micromirror device based on real-time feedback from a single-pixel infrared detector. This self-evolving process enables the autonomous reconstruction of the target's geometric profile. Simultaneously, it projects a co-modulated visible beam onto the object itself or an adjacent screen, rendering the infrared target directly perceptible to the naked eye in real-time. SIVIS achieves sensing and projection without latency under an ultra-low infrared detection limit of ~0.11 photons per pixel per frame (sub-pW/cm$^2$ level) benefited from the high sensitivity. We demonstrate SIVIS in both transmissive and reflective configurations across static and dynamic scenarios. Furthermore, we also validate SIVIS to decrypt infrared-encoded anti-counterfeiting features and visualize vascular-like structures embedded within biological tissues, leveraging the in-depth imaging ability of SIVIS. This photon-feedback-driven artificial vision framework offers a scalable and adaptive solution for ultrasensitive infrared vision, opening promising avenues for night vision, biomedical imaging, and sensing under extreme low-light conditions.

## 1. Introduction

The spectral sensitivity of the human retina is intrinsically confined to the visible (VIS) regime (400–700 nm), rendering the information-rich infrared (IR) spectrum imperceptible to the naked eye. Despite its invisibility, the IR band carries critical thermal and chemical signatures, fueling a surge of interest in fields ranging from night vision (*1*, *2*) and semiconductor inspection (*3*) to non-invasive biomedical imaging (*4*, *5*), along with a broad range of security and sensing (*6–10*). However, conventional IR cameras remain relatively costly due to the sophisticated fabrication processes required to hybridize the narrow-bandgap photodetector array (e.g., InGaAs) with the silicon readout integrated circuits (ROIC) (*11*). Furthermore, the necessity for active thermo-electric cooling to suppress dark noise leads to bulky architectures, severely hindering their miniaturization and integration into portable or wearable platforms (*12*).

Upconverting low-energy IR photons to high-energy VIS light offers an alternative approach for IR visualization, enabling direct detection by the naked eye or standard silicon-based sensors. Compared with conventional IR cameras, infrared-to-visible upconversion is a cost-effective solution with reduced noise for pixel-free IR imaging, which has attracted lots of interest in recent decades (*13*). Infrared-to-visible upconversion mechanisms can be broadly classified into two categories: nonlinear and linear optical processes (*14*). Among nonlinear methods, parametric upconversion imaging (sum-frequency generation, SFG) that can convert IR photons to VIS light by mixing the pump beam has demonstrated strong potential for IR vision and mid-IR spectroscopy (*15*). While effective, the systems typically require bulky nonlinear crystals, expensive high-power pulsed laser/intense excitation power (~1 MW/ $cm^2$), and strict phase matching, which limit their practicality in portable or low-light scenarios (*16*, *17*) Another nonlinear approach, triple–triplet annihilation based infrared-to-visible molecular photon upconversion (TTA-UC) has been investigated widely because it can operate over tunable, broadband spectral ranges and function effectively under modest excitation intensities (~1 mW·$cm^2$) (*18*). Nevertheless, TTA-UC systems often require deoxygenated organic solvents, leading to fabrication complexity and limited compatibility with *vivo* environments (*19*). In addition, rare-earth-doped upconversion nanoparticles have been widely employed in infrared upconversion imaging and bio-sensing (*20*), but they are fundamentally limited by low quantum yields and the need for relatively high excitation power (*21–24*).

On the other hand, infrared-to-visible conversion through linear photon upconversion which integrates IR-absorption photodetectors (PD) with light-emitting diodes (LED) provides an optional scheme through light–electricity–light conversion (*25*, *26*). The type of optical upconverters exhibits efficient advantages in flexible fabrication, low cost, relatively low excitation power (~1 $\mu W \cdot cm^2$), and naked-eye visibility thanks to free precise pixelization, back-to-back PD–LED structure (*27*). A variety of linear upconversion implementations have been explored, including all-inorganic (*28*), all-organic (*29*), hybrid organic–inorganic (*30*), and colloidal quantum dot (CQD)-based devices (*31*). However, despite these advances, linear photon upconversion still suffers from high excitation density, limited quantum efficiency, high turn-on voltage, and narrow detection band (*13*). Most importantly, current infrared-to-visible upconversion systems struggle to operate under ultra-low IR photon flux—especially in the single- or sub-photon-per-pixel regime—which is critical for emerging applications in quantum sensing, low-light biomedical imaging, and autonomous navigation in dark or obscured environments (*32*). This highlights the urgent need for novel upconversion approaches that combine high sensitivity, broad spectral response, device flexibility, and real-time visualization under minimal illumination.

Here, we report a novel infrared-to-visible upconversion and visualization strategy based on self-evolving infrared-to-visible upconversion with single-pixel detection (SIVIS), which enables broadband, scalable, real-time upconversion and sensing under ultra-low radiation. SIVIS optimizes digital micromirror device (DMD) patterns towards the IR target according to feedback from an IR single-pixel detector (SPD) by leveraging the genetic algorithm. Simultaneously, visible-light patterns co-modulated with the IR illumination are projected to indicate the IR scenario, enabling direct observation by the naked eye or conventional cameras. Notably, this process occurs without any perceptible delay between imaging and projection. SIVIS supports effective infrared-to-visible upconversion under IR flux as low as ~0.11 photons per pixel per detection (~0.1 pW/cm$^2$), benefiting from single-element sensing. We demonstrate its applicability in both reflective and transmissive imaging modalities, allowing real-time visualization of static and dynamic IR scenes. To validate the capability of SIVIS for infrared-to-visible applications, SIVIS has been applied to IR anti-counterfeiting using IR-sensitive security materials and to deep-tissue visualization through biological tissue, taking advantage of the long penetration depth of IR light. The proposed SIVIS offers a compact, low-cost, and scalable platform for next-generation IR visualization under extreme photon conditions, with promising applications in bioimaging, nighttime robotics, autonomous driving, and ultra-low-light surveillance, bridging the gap between invisible IR information and direct human perception/standard silicon-based sensors.

## 2. Results

The conceptual framework of the proposed SIVIS is illustrated in Fig. 1a, in which the imperceptible IR target is visualized by projecting visible patterns that mirror the target's geometry onto either the target surface or a display screen, enabling direct observation by the naked eye or a standard silicon-based camera. Unlike conventional infrared imaging based on focal-plane arrays that rely on direct image formation, SIVIS adopts real-time adaptive structured projection scheme to enable infrared-to-visible upconversion and direct visualization with a single-pixel detection. As shown in Fig. 1b, the DMD is used to spatially modulate the projection patterns onto the target, similar to the configuration used in single-pixel imaging (SPI) (*33–37*). Unlike SPI, the illumination patterns in SIVIS are not pre-designed, but would be iteratively evolved towards the target's geometry, by enhancing a designed cost function (CF) (*34*) with the real-time feedback from the IR SPD (More details in Supplementary Section S1). The CF is defined as the ratio between the detected SPD intensity and the sum of the corresponding DMD pattern. As the CF is maximized through the genetic algorithm, the illumination patterns progressively evolve toward the shape of the target. In this architecture, infrared and visible beams share a common optical path and are simultaneously modulated by the DMD. The modulated IR photons interact with the target (in either transmission or reflection configuration), and the resulting signal is collected by the IR SPD to adaptively refine the DMD patterns. Concurrently, the co-modulated visible patterns are projected onto the target or a screen, enabling the evolving IR profile to be visualized in real time by the naked eye or a standard silicon-based camera. Figure 1b illustrates this optimization trajectory, showing the transition from an initially random pattern to a well-defined "butterfly" target (Fig. 1b, lower left), with a corresponding increase in CF across successive generations (Fig. 1b, lower right). By leveraging the superior penetration depth and the exceptional detection sensitivity, SIVIS would facilitate versatile advanced applications such as IR anti-counterfeiting, IR light-guided biomedical indication, imaging or sensing in photon-starved regimes (Fig. 1c).

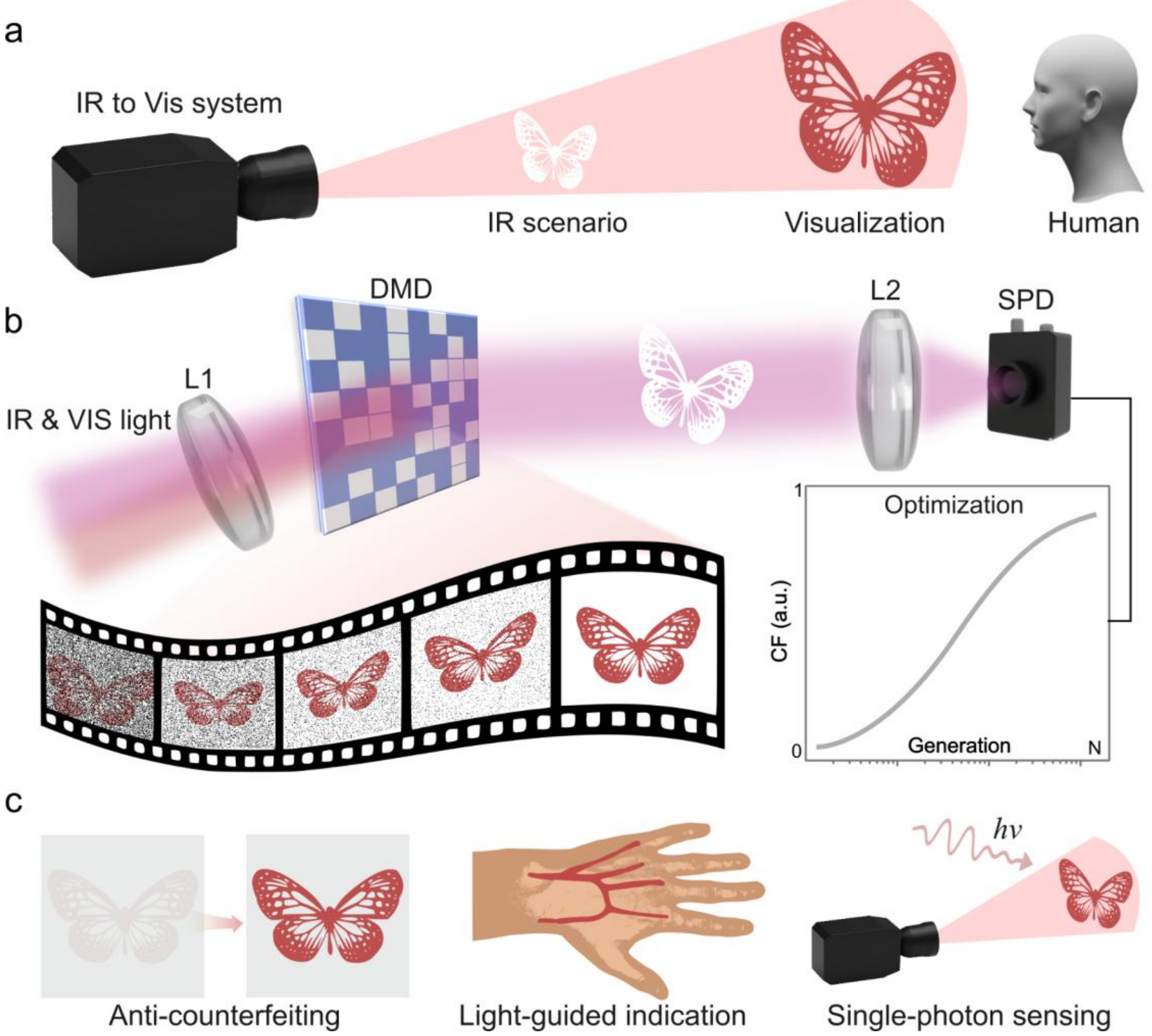

**Figure 1**. Visualizing the infrared detection via self-evolving infrared-to-visible upconversion

with single-pixel detection (SIVIS). **(a)** Illustration of SIVIS. SIVIS senses infrared targets and projects corresponding visible patterns, enabling direct observation by the naked eye or standard silicon-based cameras. **(b)** Schematic diagram of SIVIS. The IR (lavender) and VIS (pale pink) beams are spatially modulated by a single DMD concurrently. The projected patterns from the DMD are iteratively optimized toward the IR target using an optimization algorithm, with feedback from the IR SPD. In parallel, the corresponding visible patterns are projected onto the target or a screen for real-time visualization, and can be directly observed by the naked eye or a standard camera. As the optimization proceeds, the clarity of the projected visible illumination patterns improves (bottom-left) with increasing generation index (bottom-right). Meanwhile, the CF, serving as the feedback source for the evolutionary process, increases monotonically with the increased generation number. L1–L2, lenses; DMD, digital micromirror device; IR SPD, infrared single-pixel detector. **(c)** Potential applications of SIVIS for anti-counterfeiting, IR-light-guided indication, and sensing under ultra-weak light conditions, by leveraging the invisibility and deep penetration properties of IR light together with the ultrasensitive detection capability.

To validate SIVIS, we first demonstrate infrared-to-visible upconversion in both transmission and reflection configurations. The transmission setup is illustrated in Fig. 2a (see reflection setup in Fig. 3a). The near-IR (852 nm, red) and VIS (532 nm, green) beams are coaxially combined via a long-pass dichroic mirror and co-modulated by a DMD. The modulated visible patterns are diverted toward a display screen by a beam splitter, whereas the IR signal photons, after traversing the target, are collected by a lens, filtered by a band-pass (BP) filter (850/50 nm), and ultimately detected by the SPD. By maximizing the CF, the DMD patterns iteratively converge toward the target's spatial profile; simultaneously, the corresponding upconverted visible patterns are projected onto the screen, enabling real-time observation by the naked eye or a standard silicon-based camera. The DMD patterns are magnified and projected onto the object using a single projection lens, achieving a total magnification of ~23× (Supplementary Fig. S2). The illuminated field of view (FOV) is approximately 10×10 $mm^2$, corresponding to an illumination pattern of 32 × 32 pixels. The transmission and reflection targets consist of a specific region from a USAF-1951 resolution test chart and a triangular piece of white paper, respectively (Figs. 2b and 2c).

Figure 2d presents representative visible frames captured at 1 *s*, 8 *s*, and 59 *s*, by a conventional visible camera, during the optimization process for the static transmission target. As evolution progresses, the shape of the target begins to emerge at 1 *s*. The final frame exhibits a high-fidelity, background-free visualization, demonstrating the system's potential for high-quality infrared-to-visible upconversion. The cross-sectional intensity profiles along the marked white line (the average intensity of the central ~10 pixels) are shown in the right panel, while the corresponding DMD patterns are shown in the lower right. The profiles and patterns exhibit a consistent trend of improved clarity with increasing generations. We further evaluated SIVIS in a dynamic scenario with the target translating at 0.1 mm/s. Figure 2e shows selected frames at 28 *s*, 60 *s* and 88 *s*, representing entering, staying, and leaving phases, respectively (See Visualization 1). The reflection modality is similarly demonstrated in Figs. 2f and 2g, corresponding to the static and dynamic conditions, respectively. Figure 2h plots the CF and contrast-to-noise ratio (CNR) curves for the static transmission target in Fig. 2d, both metrics increase steadily with successive generations, indicating progressive optimization. The corresponding performance metrics for the dynamic case are shown in Fig. 2i. For the dynamic case, both CF and CNR curves exhibit a characteristic rise-and-fall profile, peaking when the target fully occupies the field of view (FOV), as highlighted by the light blue and green shaded regions. These results confirm SIVIS ability to track and upconvert both static and dynamic IR scenes with high contrast and minimal background noise.

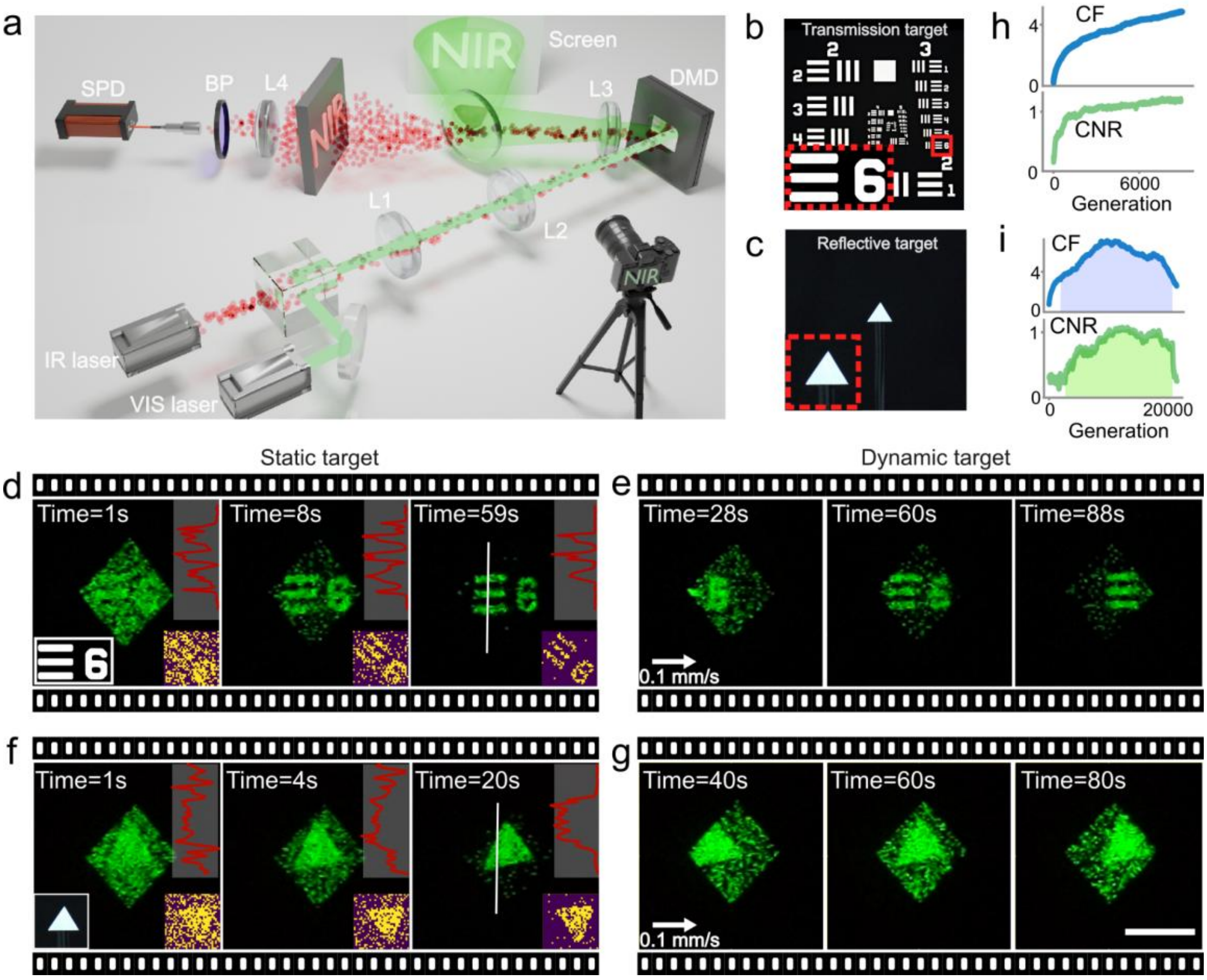

**Figure 2.** Demonstration of SIVIS with static and dynamic targets in transmission/reflection mode. **(a)** Schematic diagram of SIVIS in transmissive configuration. The infrared (IR) and visible (VIS) beams are combined by a dichroic mirror (DM), and incident on the DMD simultaneously. The modulated patterns are projected onto the target with a projection lens. Transmitted IR photons are collected by a convex lens, spectrally filtered by a band-pass filter (BP), and subsequently detected by the SPD. Meanwhile, the corresponding visible patterns are deflected to the target or the screen. **(b-c)** The transmission and reflection targets. **(d-e)** Temporal evolution of the upconverted visible projection patterns for the static and dynamic transmission targets. **(f-g)** Representative frames of the upconverted visible projection patterns featuring static and dynamic triangular targets in a reflective configuration. **(h)** Evolution of CF and CNR as a function of generation for the static transmission target in (d). **(i)** Evolution of CF and CNR as a function of generation for the dynamic transmission target in (e). The scale bar is 10 mm.

IR vision has been widely utilized in anti-counterfeiting, leveraging the inherent imperceptibility of IR-active features under standard visual inspection. Here, we demonstrate SIVIS for anti-counterfeiting, where invisible IR security features are rendered perceptible by directly projecting their visible counterparts onto the target surface. As illustrated in Fig. 3a, both IR and visible illumination patterns are directly projected onto the target (e.g., a "cat paw" shape). The target's spatial profile is subsequently rendered visible through the self-evolved patterns, enabling direct observation by the naked eye. Here, the paper-based targets with various shapes (e.g., a "cat paw", "N", "I", "R", and "123") are fabricated using IR-absorbing security inks that remain imperceptible under ambient lighting. Figure 3b shows the photograph of an example target captured by a conventional camera, where the lower-left inset highlights the inked regions (demarcated by white circles) that are otherwise indistinguishable from the background. The upper right inset shows the IR powder dissolved in anhydrous ethanol, while the right panel plots its absorption spectrum with an absorption peak at ~852 nm.

We further investigate SIVIS in both negative and positive visualization modes. The negative mode is implemented by simply taking the reciprocal of CF (Supplementary Section S1). Figure 3c and 3d present the dual-mode visualization results, where the projected visible images exhibit spatial and intensity complementarity, as a direct consequence of the inverted DMD patterns. The right panels plot the CF evolution along with the final and time-averaged DMD patterns. While the CF curves exhibit similar convergence profiles, the final projections are inverted in appearance due to the reciprocal formulation. Representative frames from the visualization videos (0–16 s; Visualization 2) in the lower panels illustrate the progressive emergence of the hidden IR features. These dual-mode results are consistently observed across other static targets ("N," "I," "R") as shown in Fig. 3e. Finally, the dynamic performance of SIVIS is evaluated by moving the target with the handwritten digits ("123") at 0.1 mm/s (Fig. 3f and Visualization 3). The contours of the digits are marked with yellow dashed lines, shown in both negative and positive modes. SIVIS progressively evolves the illumination patterns toward the target shape, resulting in spatially concentrated illumination that is predominantly focused on the object of interest. By leveraging reversible DMD illumination, SIVIS not only enables direct, real-time visualization of IR-hidden content without reconstruction, but also improves photon efficiency compared to conventional single-pixel imaging followed by post-projection, where the entire field of view is illuminated and the image is reconstructed computationally. This intrinsic coupling between imaging and projection further suggests transformative potential in mitigating photobleaching and photodamage in sensitive materials, as well as enabling spatially selective photothermal therapy (*38*).

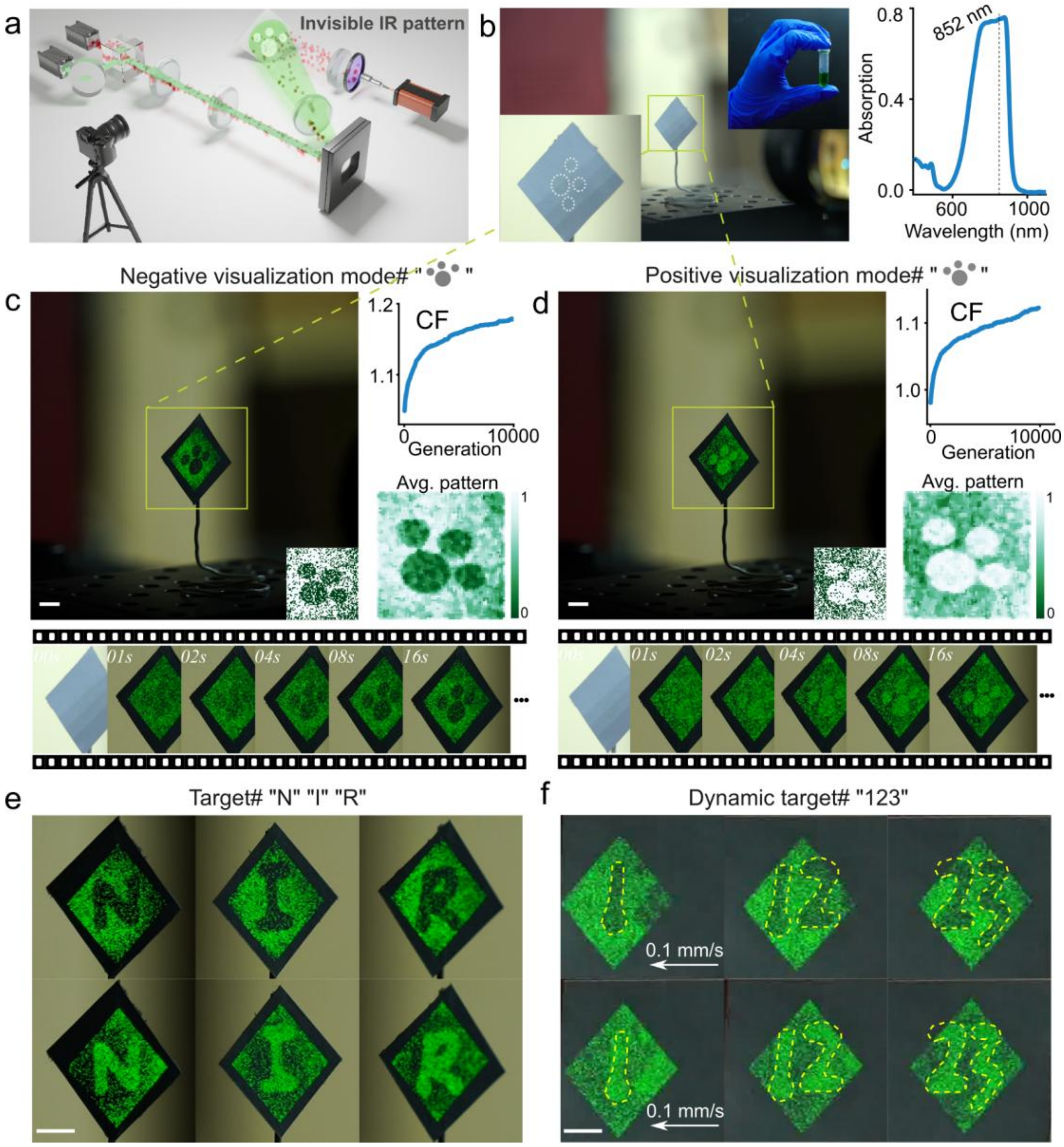

**Figure 3**. In-situ infrared-to-visible upconversion visualization of SIVIS in reflective configuration featuring positive and negative modes. **(a)** Schematic illustration of SIVIS in reflective modality. Infrared targets are fabricated using IR-absorbing ink in different shapes, such as a “cat paw” or numbers, which are invisible. SIVIS can visualize specific patterns in situ, allowing the information to be directly perceived by the naked eye. A standard commercial camera is employed to capture the upconverted visible projection patterns. **(b)** Photograph of the cat paw target printed on an A4 paper. The lower left inset provides an enlarged view where the inked regions (white circles) remain imperceptible under visible light. The upper right displays the ink dissolved in anhydrous ethanol, while the right panel presents the corresponding absorption spectrum, exhibiting a peak at ~852 nm. **(c)** Visualization of the cat paw target with SIVIS in negative mode. The left visible photograph, captured under ambient light, uses a yellow box to indicate the field of view corresponding to (b). The upper-right panel plots CF curve as a function of generation. The middle-right insets display the final generation pattern and the averaged pattern across the evolution. The bottom panel presents selected frames extracted from the visualization projections at different time points. **(d)** Visualization result of the cat paw target under SIVIS positive mode. **(e)** Visualization results of “N”, “I”, and “R” under negative (top) and positive (bottom) modes. **(f)** Dual-mode visualization of the dynamic “123” target with a lateral speed of 0.1 mm/s. The yellow dashed lines highlight the targets’ boundaries for clarity. The scale bar is 10 mm.

We further demonstrate that SIVIS is capable of subsurface feature conversion imaging, enabled by its NIR imaging capability and highly sensitive detection strategy. SIVIS was employed for near-infrared (NIR) light-guided indication to visualize subsurface vascular structures within a hand phantom (Fig. 4a). The phantom was constructed by patterning vein-like structures with IR-absorbing ink onto a wooden hand model, which was subsequently covered with ~1.5 mm-thick layer of uniform artificial skin (Fig. 4a, upper-right inset). As shown in Fig. 4b, the vascular features remain imperceptible under ambient lighting. SIVIS was then leveraged to upconvert the hidden IR information into the visible domain. Figure 4c presents representative frames from the dual-mode visualization sequence at 0 $s$, 10 $s$, 20 $s$, 40 $s$, and 70 $s$ (see Visualization 4).The initial frame at $t = 0$ s presents the vessel’s location with yellow solid lines in the lower-left panel. The subsequently upconverted visible projection patterns undergo temporal evolution and reveal the vascular structure with increasing clarity. Although the captured images exhibit a degree of blurring compared to non-scattering targets shown above, the vascular structures remain clearly distinguishable to the naked eye. Furthermore, we demonstrate SIVIS through biological tissue, as shown in Fig. 4d. The "T"-shape target on a microscope slide is hidden by the chicken tissue with a thickness of ~2 mm, highlighted by the yellow dash line, as shown in Fig. 4e. Figure 4f shows the final visible output from SIVIS, enabling the embedded target visible on the screen under ambient conditions. These results validate the potential of SIVIS for in-situ visualization through biological media, highlighting its utility in non-invasive imaging under scattering conditions and light-guided indications in medical applications.

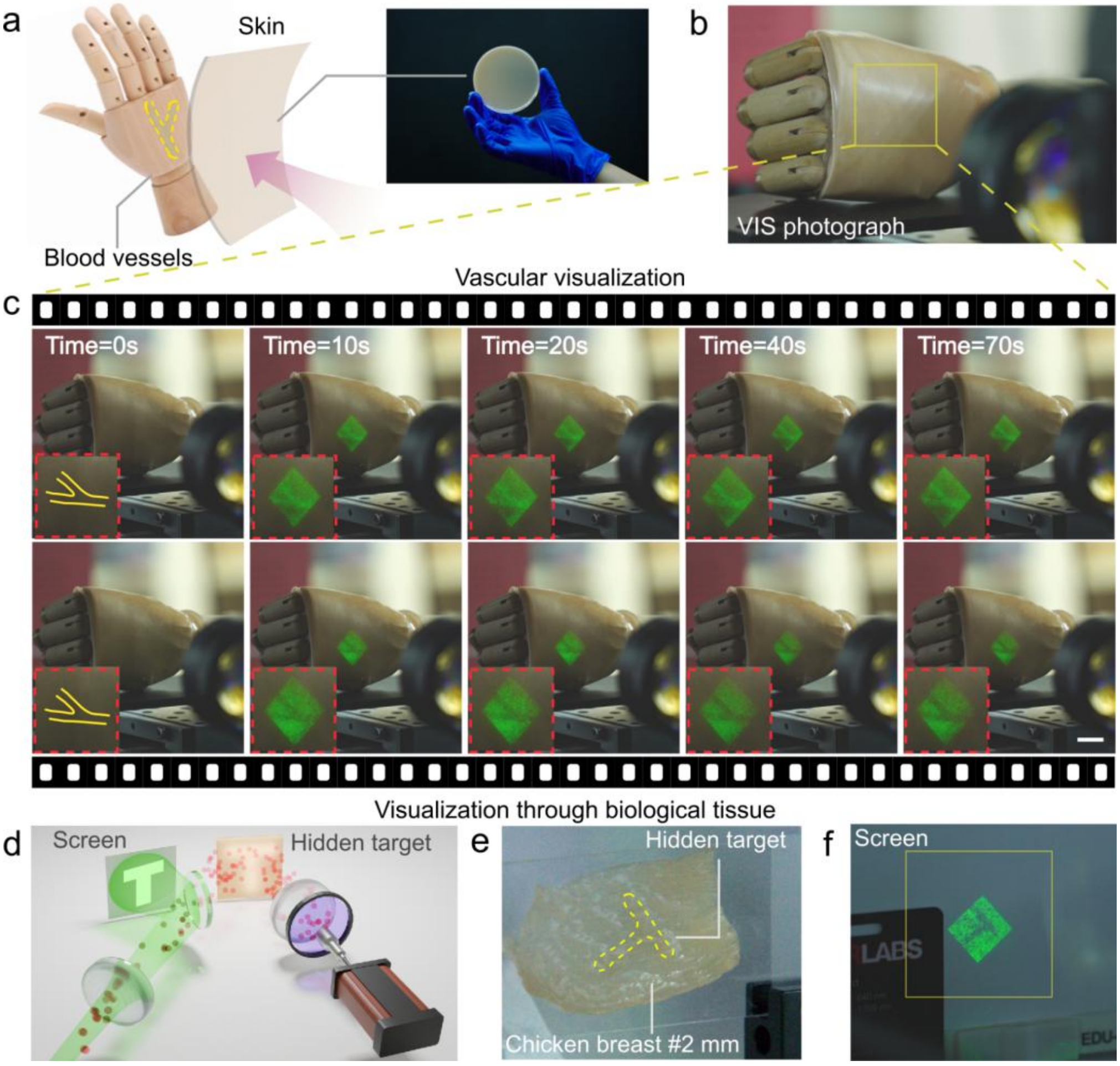

**Figure 4**. NIR light-guided indication via SIVIS. **(a)** Schematic illustration of the tissue phantom, featuring a vein-like pattern fabricated with infrared-absorbing ink to mimic subsurface vasculature beneath artificial skin. The upper-right inset presents a photograph of the artificial skin, which serves as the scattering medium. **(b)** Photograph of the hand phantom, comprising a wooden hand model integrated with artificial skin and the vein-mimicking IR pattern. **(c)** Selected frames from positive-negative SIVIS visualization of the vein-mimicking structure. The insets provide the enlarged regions of interest. The upconverted visible projection patterns are captured directly by a standard commercial camera. **(d)** Schematic illustration of visualizing a tissue-hidden target using SIVIS. **(e)** Photograph of the "T"-shaped target covered by a ~2 mm-thick chicken breast on a microscope slide. **(f)** Final visualization result of the target on the screen via SIVIS. The scale bar is 10 mm.

SIVIS enables infrared-to-visible upconversion under ultralow IR flux conditions, reaching the single- or even sub-photon level, as shown in Fig. 5a, benefiting from ultrahigh photon sensitivity of single-pixel single photon detectors. We evaluated the performance of SIVIS across five flux levels, with average photon counts of 0.11, 0.38, 1.30, 3.86, and 12.69 photons per pixel (ppp) per detection frame, achieved by varying the acquisition time from 0.05, 0.15, 0.5, 1.5 to 5 *ms*. As shown in Fig. 5b, the CF increases steadily with successive generations across all flux levels. At the higher flux (12.69 ppp), the system exhibits rapid CF improvement due to the superior signal-to-noise ratio (SNR). In contrast, in the photon-starved conditions (0.11 and 0.38 ppp), although CF evolution is slower, it maintains a clear upward trend. The inset highlights the CF curve at 0.11 ppp, confirming SIVIS's robust operability under extremely low-light conditions. In the meantime, the detected photon counts also increase as the generation progresses at all flux levels. Figure. 5c presents the photon counts versus generation for the 0.11 and 3.86 ppp conditions. At 12.69 ppp, SIVIS produces a high-quality, background-free visible illumination pattern corresponding to the transmissive "three-stripe" target (Fig. 5d and Visualization 5). Figure 5e presents the CNR values of the final visualization images as a function of IR photon flux. As expected, the CNR improves monotonically with increasing flux. The small inset figures of Fig. 5e show the final visualization results and corresponding average DMD patterns (marked by magenta boxes), highlighting the clear quality enhancement at higher flux levels. Notably, even at 0.11 ppp, the system achieves discernible target visualization, with clarity progressively improving as flux increases. At 0.11 and 12.69 ppp, the photon distribution histograms throughout the evolution process follow the Gaussian distribution from 0.08-0.16, and 10.00-16.00 ppp, respectively (Fig. 5e). We also demonstrate the tracking capability of SIVIS with a moving target under ~1.00 ppp (Supplementary Fig. S6). These results demonstrate the capability of SIVIS to operate in sub-photon flux regimes, providing robust upconversion and visualization for ultrasensitive imaging and sensing. To highlight this, we compare the IR detection intensity of SIVIS with state-of-the-art upconversion methods developed over the past decade (Fig. 5f). The nonlinear upconversion approaches are highlighted in light green and linear approaches are highlighted in pink, respectively. The nonlinear optical SFG upconversion has high excitation power at ~1 MW/ $cm^2$ due to the need for the pulsed laser (*16*, *39*). Most upconversion approaches have a relatively high excitation power at ~$10^2$ to $10^4$ μW/$cm^2$ level (*11*, *12*, *18*, *23*, *25*, *26*, *29*, *31*, *40*). In contrast, SIVIS maintains functionality at an ultra-low IR detection power of ~0.1 pW/$cm^2$ level, representing a reduction in power requirements by ~8 orders of magnitude compared with existing methods (Supplementary Section S4).

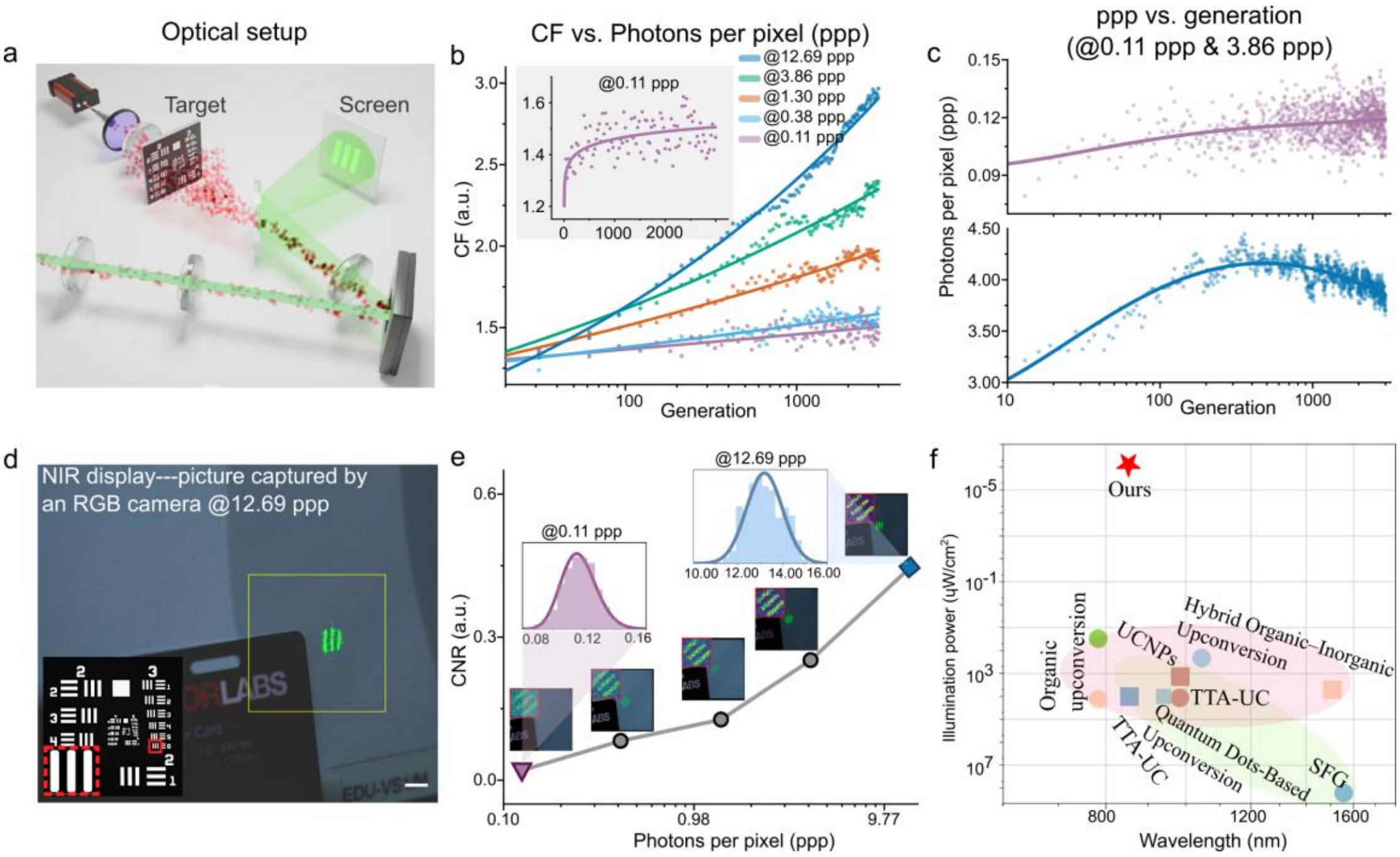

**Figure 5**. Sub-photon flux SIVIS. **(a)** Schematic illustration of SIVIS in the single-photon condition. **(b)** CF evolution under different average photons per pixel (ppp) levels, ranging from 12.69 to 0.11 ppp. The inset shows a magnified view of CF evolution at the extreme low-light level of 0.11 ppp. **(c)** Detected photons per pixel as a function of the generation for the 0.11 ppp (top) and 3.86 ppp (bottom) levels. **(d)** SIVIS visualization of the IR target at 12.69 ppp under ambient light. The inset illustrates the target—a specific region of the USAF-1951 resolution test chart. Scale bar: 10 mm. **(e)** CNR of the captured upconversion images versus ppp across the five different photon levels. The insets showcase the upconversion images and average evolved DMD patterns (marked with purple boxes) throughout the optimization process. The two upper insets present photon distribution histograms at 12.69 and 0.11 ppp levels, illustrating the photon statistics across the evolution. **(f)** Comparison of the state-of-the-art upconversion methods. The linear and non-linear infrared-to-visible upconversion methods are highlighted in pink and green, respectively. The main upconversion approaches require illumination power at ~$10^3$ μW/$cm^2$ level, while SIVIS remains at an ultra-low detection power of ~0.1 pW/$cm^2$ level.

## 3. Discussion

In summary, we have developed SIVIS, a self-evolving infrared-to-visible protype, to upconvert IR vision to visible light projections, based on the single-pixel detection feedback and evolved illumination patterns. In contrast to existing infrared-to-visible methods, which face tremendous challenges in photon-starving conditions due to the intense IR excitation intensity, SIVIS can facilitate high-fidelity visualization under sub-photon flux irradiation (~0.11 photons per pixel per detection), owing to the availability of highly sensitive and broadband detection with a single-pixel photodetector. We validated SIVIS in both transmission and reflection configurations, demonstrating reliable performance for static and dynamic scenes. We also demonstrate SIVIS for IR anti-counterfeiting and biological tissue visualization, showcasing the potential in high-contrast optical visualization and non-invasive medical diagnostics.

We note that the current limitation of SIVIS lies primarily in the insufficient optimization efficiency of the conventional genetic algorithm, resulting in an infrared-to-visible upconversion time of ~0.5 s from random speckle to recognizable shape. Future work can address this challenge by incorporating faster optimization algorithms, such as gradient descent or reinforcement learning, and by improving the hardware throughput—for example, increasing the data transfer bandwidth between the computer and the DMD. While SIVIS currently operates using a dual-wavelength, active structured illumination scheme, it can also be adapted to a passive configuration.

Tailored for specific application scenarios, SIVIS also supports dual-mode (positive and negative) visualization/illumination by switching between the CF and its reciprocal. This framework offers a flexible modality that either highlights IR-sensitive/salient regions (positive mode) or minimizes photo degradation of IR photosensitive materials (negative mode), potentially enabling the targeted treatment of diseased tissues while ensuring the maximal preservation of surrounding healthy structures (*38*). SIVIS also offers a promising pathway for next-generation surgical guidance by integrating with augmented/virtual reality (AR/VR) enabling the real-time overlay of invisible IR-coded physiological information onto a visible display under low-light conditions (*41*, *42*). Furthermore, the single-pixel architecture of SIVIS offers a compact and cost-effective pathway for achieving wide-spectrum conversion from X-ray to terahertz waves (*33*, *43*). SIVIS provides a novel, scalable, and photon-efficient solution for infrared-to-visible upconversion with superior sensitivity. We, therefore, anticipate that this work would open up new dimensions for high-sensitivity infrared visualization, offering a versatile platform for applications ranging from night vision and IR sensing to non-invasive imaging and light-guided biomedical applications.

## 4. Methods

### 4.1 Experimental setup

All experiments were conducted with the purpose-built SIVIS (Fig. 2a and Supplementary Fig. S2). The IR and VIS lasers are a polarization-maintaining single-mode continuous-wave 852 nm laser (MDL-III-852L, CNI) and a free-space output 532 nm laser (MGL-FN-532-500 Mw, CNI). The collimated IR and VIS laser beams have diameters of 10 and 20 mm, respectively. A long-pass dichroic mirror is used to combine the two laser beams and then co-modulated by a DMD (GmbH V-7001, ViALUX). To accelerate the DMD loading process, only the central 32/64 rows of DMD are utilized, and the DMD operates at a 10 kHz refresh rate, enabling a time interval of ~5 ms with 10 patterns per generation, corresponding to 200 generations per second. The patterns are projected and magnified by a single $f$ = 50 mm projection lens which is placed at ~55 mm from the DMD. The distance between the projection lens and the target is ~0.5 m. A microscope glass slide serves as the BS to part the VIS light for display. The linear motorized translation stage (EM-LSS90-200C1, LBTEK) can carry the target with a constant speed. Then a $f$ = 100 mm convex lens collects the transmitted/reflected light which then is guided by a 200 μm core diameter (FCM2-PC-200L, JCOPTIX) to SPD in single-photon experiments. Two types of SPD are utilized in experiments. The one used in Figs. 2-4 is an analog detector (Thorlabs, PDA100A2). The other is a digital photon counter (Excelitas, SPCM- AQRH-14) with a ~50% photon detection efficiency at 852 nm and a 180-μm sensor diameter. A commercial camera (Sony Alpha 7 III) is employed to record visible pictures and videos with an exposure time of 40 to 500 ms and an aperture of f/5.6. The SPD and DMD are controlled and synchronized using a multifunction data acquisition device (USB-6353, National Instruments) and a purpose-built Python program.

### 4.2 Optimization

The cost function was optimized by a genetic algorithm with a population size of 10 for Figs. 1 to 5. The initial and final mutation rates are $R_0$ = 0.01 and $R_{end}$ = 0.0025, respectively. A varying ratio between the increasing generation index ($g$) and the fixed total generation ($m$) serves as the decay factor, while the mutation rate is defined as follows:

$$r = (R_0 - R_{end}) \cdot e^{-\frac{g}{m}} + R_{end}, \quad (1)$$

where the mutation rate decays exponentially. Increasing the population size can enhance evolutionary efficiency at the cost of longer loading time. Therefore, we selected a small population of 10 to balance the trade-off. The initial patterns are randomly generated and sorted according to the CFs. Offspring patterns are then derived from parent patterns through probabilistic selection and crossover. To ensure that higher-ranking patterns have a greater chance of selection, we apply a cumulative summation to an ascending sequence [1, 2, 3, … $k$], producing [1, 3, 6, …$\frac{k(k+1)}{2}$]. A random number is drawn from the interval [1 $\frac{k(k+1)}{2}$]; the index corresponding to the interval in which the number falls determines the selected parent pattern. This selection scheme assigns higher probabilities to patterns with high CFs. Offspring are generated by randomly mixing the pixels from two selected patterns followed by a mutation step. Mutation is implemented by flipping a number of pixel states (from 0 to 1 or 1 to 0), proportional to the product of the total number of pixels and the mutation rate. The newly generated offspring will replace the lower half of the population for the next generation.

**Supplementary Information**

Supplementary Information is available from the author.

**Acknowledgments**

This work was supported by the National Natural Science Foundation of China (U23A20481, 62503032, 62275010, 62573029). We are grateful to the Atomic-Scale In Situ Fabrication Platform of the Analysis & Testing Center of Beihang University for the facilities, and the scientific and technical assistance.

**Conflict of interest**

The authors declare that they have no conflict of interest.

**Data Availability Statement**

The data that support the findings of this study are available from the corresponding author upon reasonable request.